\documentstyle[aps,epsf,preprint]{revtex}
\tightenlines
\begin{document}
\draft
\title{On the masses, branching ratios, and full widths of heavy $\rho^\prime$,
$\rho^{\prime\prime}$ and $\omega^\prime$, $\omega^{\prime\prime}$
resonances.}
\author{N.~N.~Achasov \footnote{Electronic address: achasov@math.nsc.ru}
and
A.~A.~Kozhevnikov \footnote{Electronic address: kozhev@math.nsc.ru}}
\address{Laboratory of Theoretical Physics, \\
Sobolev Institute for Mathematics \\
630090, Novosibirsk-90, Russia}
\date{\today}
\maketitle
\widetext
\begin{abstract}
Based on the previous and recent  fits of multiple data on the
reactions of $e^+e^-$ annihilation, $\tau$ lepton decay, and the
reaction $K^-p\to\pi^+\pi^-\Lambda$, the magnitude of the
branching ratios and  total widths of the isovector $\rho^\prime$,
$\rho^{\prime\prime}$, and the isoscalar $\omega^\prime$,
$\omega^{\prime\prime}$ resonances are calculated. Some topics on
the spectroscopy of the $\rho(1450)$ and $\omega(1420)$ states are
discussed.
\end{abstract}
\pacs{PACS number(s):13.65.+i, 13.25 Jx, 14.40.Cs}
\narrowtext

The situation with the resonances
$\rho^\prime\equiv\rho^\prime_1\equiv \rho(1450)$,
$\rho^{\prime\prime}\equiv\rho^\prime_2\equiv \rho(1700)$, with
isospin one and those $\omega^\prime\equiv
\omega^\prime_1\equiv\omega(1420)$, $\omega^{\prime\prime}\equiv
\omega^\prime_2\equiv\omega(1600)$, with isospin zero is still far
from being clear. Although the characteristic peaks corresponding
to these resonances were observed in a number of channels of the
one-photon $e^+e^-$ annihilation, $\tau$ lepton decays, $N\bar N$
annihilation, photoproduction, etc., the specific masses and
branching ratios are not properly established and hence are not
given in the Tables \cite{pdg}. Recently, the present authors
undertook an attempt to fit then existing data on $e^+e^-$
annihilation, $\tau$ lepton decays, and the reaction
$K^-p\to\pi^+\pi^-\Lambda$, where the above heavy resonances were
observed, in the framework of the unified approach. The approach
is based on the scheme that takes into account both energy
dependence of the partial widths and  the mixing via common decay
modes among the latter and the ground state $\rho(770)$ and
$\omega(782)$ resonances, in the respective isovector \cite{ach97}
and isoscalar \cite{ach98} channels. When so doing, the masses and
{\it coupling constants} of the resonances were taken as free
parameters to be determined from the fit. This choice is
convenient because, first, it is just these parameters that are
essential in identifying the nature of heavy resonances with
$J^{CP}=1^{--}$. Second, the function $\chi^2$ used in obtaining
the intervals of the variation of the extracted parameters is
represented in this case as the sum of independent contributions,
so that the covariance matrix (see the statistics section in
Ref.~\cite{pdg}) is diagonal. However, the results of the
measurements are often represented in terms of the masses and {\it
branching ratios}. In view of the excessive size of the
publications \cite{ach97,ach98} we refrained at that time from the
calculation of the partial widths with the parameters found in
\cite{ach97,ach98}. Here we fill this gap and calculate the
branching ratios and total widths of the $\rho^\prime_{1,2}$ and
$\omega^\prime_{1,2}$ resonances using the new data on the cross
section of  the reactions $e^+e^-\to\pi^+\pi^-\pi^0$
\cite{cmd298,snd99}, $e^+e^-\to\omega\pi^0$ \cite{snd00}, and the
data on the spectral function $v_1$ measured in the decays
$\tau^-\to\omega\pi^-\nu_\tau$,
$\tau^-\to\pi^+\pi^-\pi^-\pi^0\nu_\tau$ \cite{cleo}. We also
comment on the issue of why the resonances with the masses greater
than 1400 MeV {\it should be} wide in the conventional $q\bar q$
model, and why their bare masses get shifted towards the greater
values as compared to the visible peak positions. The latter
point, as will be shown below,
 is of the direct relevance to the hadronic
spectroscopy of the $\rho(1300)$ resonance reported by the LASS
group \cite{lass}, and the $\omega(1200)$ resonance reported by
the SND team \cite{snd99}.

First, let us try to understand why the situation with the
resonances possessing the masses greater than 1400 MeV is so
complicated despite the numerous experiments aimed at the
determination of their masses and partial widths. As is known
\cite{pdg}, the indications on the existance of the
$\rho^\prime_{1,2}$ and $\omega^\prime_{1,2}$ resonances were
obtained, in particular, in the $VP$ channels, where $V$, $P$
standing, respectively, for the vector, pseudoscalar meson. It is
also known that the typical magnitude of the $VVP$ coupling
constant is $g_{\omega\rho\pi}\simeq14.3$ GeV $^{-1}$. All other
$VVP$ coupling constants can be expressed through
$g_{\omega\rho\pi}$ via the SU(3) Clebsh-Gordan coefficients or,
equivalently, the quark model relations. From the point of view of
the simplest $q\bar q$ quark model, there are no reasons to expect
that the $V^\prime_{1,2}VP$ coupling constants should be
drastically suppressed as compared to the $VVP$ ones
\cite{gerasimov}. Hence, taking
$g_{\rho^\prime_{1,2}\omega\pi}\sim g_{\omega^\prime_{1,2}\rho\pi}
\simeq10$ GeV $^{-1}$, and $m_{\rho^\prime_1}\approx
m_{\omega^\prime_1}=1400$ MeV, one finds
\begin{eqnarray}
\Gamma_{\rho^\prime_1\to\omega\pi}&=&g^2_{\rho^\prime_1\omega\pi}
q^3_{\omega\pi}(m_{\rho^\prime_1})/12\pi\sim280\mbox{ MeV},
\nonumber\\
\Gamma_{\omega^\prime_1\to\rho\pi}&=&g^2_{\omega^\prime_1\rho\pi}
q^3_{\rho\pi}(m_{\omega^\prime_1})/4\pi\sim820\mbox{ MeV}.
\label{wid1}
\end{eqnarray}
Here
\begin{eqnarray}
q_{bc}(m_a)&=&
\left\{\left[m^2_a-(m_b+m_c)^2\right]\right.   \nonumber\\
&&\left.\times\left[m^2_a-(m_b-m_c)^2\right]
\right\}^{1/2}/2m_a
\label{q}
\end{eqnarray}
is the momentum of the final particle $b$ or $c$ in the rest frame
system of the particle $a$, in the decay $a\to b+c$, and the three
isotopic modes in the $\omega^\prime_1\to\rho\pi$ decay are taken
into account. To appreciate the rapid growth of the $VP$ widths
with energy, their evaluation, assuming the masses 1200 MeV,
gives, respectively, 92 and 295 MeV. Analogously, assuming that
$m_{\rho^\prime_2}\approx m_{\omega^\prime_2}=1750$ MeV, one finds
\begin{eqnarray}
\Gamma_{\rho^\prime_2\to\omega\pi}&\sim&880\mbox{ MeV},  \nonumber\\
\Gamma_{\omega^\prime_2\to\rho\pi}&\sim&2600\mbox{ MeV}.
\label{wid2}
\end{eqnarray}
Since the $VP$ decay modes are not the only ones to which heavy
resonances can decay \cite{pdg}, the resonances
$\rho^\prime_{1,2}$ and $\omega^\prime_{1,2}$, in fact, {\it
should} be rather wide. The large width of a resonance is one of
the obstacles in its identification, because such resonance often
reveals itself as the rather smooth feature in the energy behavior
of the cross section.

The second obstacle, as was pointed  out in \cite{ach97,ach98}, is
the shift of the resonance peak position from the input value of
the resonance mass. Indeed, let us consider, for simplicity, the
single resonance $R$ with the bare mass $m_R$ observed in some
channel $f$ of $e^+e^-$ annihilation, $e^+e^-\to R\to f$. Then the
cross section of the above process can be written as
\begin{eqnarray}
\sigma(s)&=&12\pi m^3_R\Gamma_{Rl^+l^-}(m_R)g^2_{Rf}   \nonumber\\
&&\times{s^{-3/2}W_{Rf}(s)\over (s-m^2_R)^2+s\Gamma^2_R(s)},
\label{sigma}
\end{eqnarray}
where $s$ is the square of the total center-of-mass energy,
$\Gamma_{Rl^+l^-}(m_R)$ is the leptonic width of the resonance
evaluated at $\sqrt{s}=m_R$. The partial hadronic width of the
decay $R\to f$ is represented in the form
$$\Gamma_{Rf}(s)=g^2_{Rf}W_{Rf}(s),$$ where $g_{Rf}$ is the
coupling constant of $R$ with the final state $f$, and $W_{Rf}(s)$
being the dynamical phase space factor of the decay $R\to f$ that
includes the possible resonance intermediate states as, for
example, in the decay $\omega\to\rho\pi\to3\pi$. The total width
of the resonance is $$\Gamma_R(s)=\sum_f\Gamma_{Rf}(s).$$ The peak
position is given by the condition of the vanishing derivative of
$\sigma(s)$ with respect to $s$:
\begin{equation}
s-m^2_R={1\over G(s)}\left\{1\pm\left[1+s\Gamma^2_RF(s)
G(s)\right]^{1/2}\right\}, \label{posit}
\end{equation}
where
\begin{eqnarray}
G(s)&=&\left[\ln\left(s^{-3/2}W_{Rf}\right)\right]^\prime,
\nonumber\\
F(s)&=&\left[\ln\left(s^{5/2}\Gamma^2_R/W_{Rf}\right)\right]^\prime.
\label{fg}
\end{eqnarray}
 Hereafter,
$\Gamma_R\equiv\Gamma_R(s)$, $W_{Rf}\equiv W_{Rf}(s)$, and prime
denotes the differentiation with respect to $s$. Eq.~(\ref{posit})
should match with the usual expression $s-m^2_R=0$ in the limit of
slow varying narrow width, hence the lower minus sign should be
chosen in Eq.~(\ref{posit}). The latter is still very complicated
equation for the determination of the peak position, so the
numerical methods should be invoked for its solution. However, all
necessary qualitative conclusions can be drawn  upon approximating
the right hand side of this equation by taking its value at
$s=m^2_R$. One can convince oneself that the dominant decay modes
of heavy resonances have the phase space factors growing faster
than the decrease of the leptonic width as $s^{-3/2}$. Then the
function $G(s)$ is positive. Also positive is the function $F(s)$.
Hence, the factor following $G^{-1}(s)$ in Eq.~(\ref{posit}) is
negative. To first order in the derivative of the phase space
volume, one finds from Eq.~(\ref{posit}) the peak position $s_R$:
\begin{equation}
s_R\approx m^2_R-{1\over2}s\Gamma^2_RF(s)|_{s=m^2_R}.
\label{posita}
\end{equation}
 One can see that
in the case of the sufficiently narrow ($\Gamma_R<200$ MeV)
resonance such as $\rho(770)$, $\omega(782)$, and $\phi(1020)$
one, the peak position, with a good accuracy, coincides with the
bare mass $m_R$.

The situation  changes when the resonance is wide, as it takes
place in the case of the $\rho^\prime_{1,2}$ and the
$\omega^\prime_{1,2}$ one. See Eqs.~(\ref{wid1}) and (\ref{wid2}).
The peak position is shifted towards the lower value as compared
to the magnitude of the bare mass. This is just what was revealed
in the fits \cite{ach97,ach98}. The more the width of the
resonance width is, the more it is shifted, so that, say, the
$\rho^\prime_2$ and $\omega^\prime_2$ resonances with the bare
masses around 1900 MeV are revealed as the peaks at 1500 $-$ 1600
MeV.

After these preliminary remarks let us present the results of the
calculation of the branching ratios and full widths. As in Refs.~
\cite{ach97,ach98}, the results are given for each channel where
the indication on the specific resonance exists. The procedure
presented here is as follows. First, we calculate the partial
widths and their errors from the errors of the masses and coupling
constants found in \cite{ach97,ach98}. Second, upon dividing each
of these partial widths by the sum of their central values, the
branching ratios and their errors are calculated. The leptonic
widths of the $\rho^\prime_{1,2}$ are taken from \cite{ach97},
while those of $\omega^\prime_{1,2}$ are calculated from the
extracted masses and leptonic coupling constants
$f_{\omega^\prime_{1,2}}$ in \cite{ach98} according to the
expression
$$\Gamma_{\omega^\prime_{1,2}}={4\pi\alpha^2m_{\omega^\prime_{1,2}}
\over3f^2_{\omega^\prime_{1,2}}}.$$ Furthermore, as compared to
Ref. \cite{ach97,ach98},  the recent data on the reactions
$e^+e^-\to\pi^+\pi^-\pi^0$ \cite{cmd298,snd99},
$e^+e^-\to\omega\pi^0$ \cite{snd00}, and $\tau$ lepton decays
\cite{cleo} are used to extract the necessary resonance
parameters. The results of the evaluation of the branching ratios
and full widths are shown in Tables \ref{tab1}, \ref{tab2},
\ref{tab3}. One can see that the simple qualitative estimates
displayed in Eqs.~(\ref{wid1}) and (\ref{wid2}), assuming the
modes besides the $VP$ one are included, agree with the results
presented there.

Now, some remarks on the spectroscopy of the heavier vector mesons
are in order. First, the $\rho(1300)$ state reported by the LASS
detector team \cite{lass} who studied the reaction
$K^-p\to\pi^+\pi^-\Lambda$,  revived an old discussion concerning
the possible existence of the $\rho(1250)$ meson, in addition to
the $\rho(1450)$ claimed to be  observed in $e^+e^-$ annihilation.
The results presented in Table \ref{tab1} show that the
corresponding peak observed by the LASS group should be attributed
to the same state $\rho(1450)$ as that presented in Reviews of
Particle Physics (RPP)\cite{pdg}.

\begin{figure}
\centerline {\epsfysize=7in \epsfbox{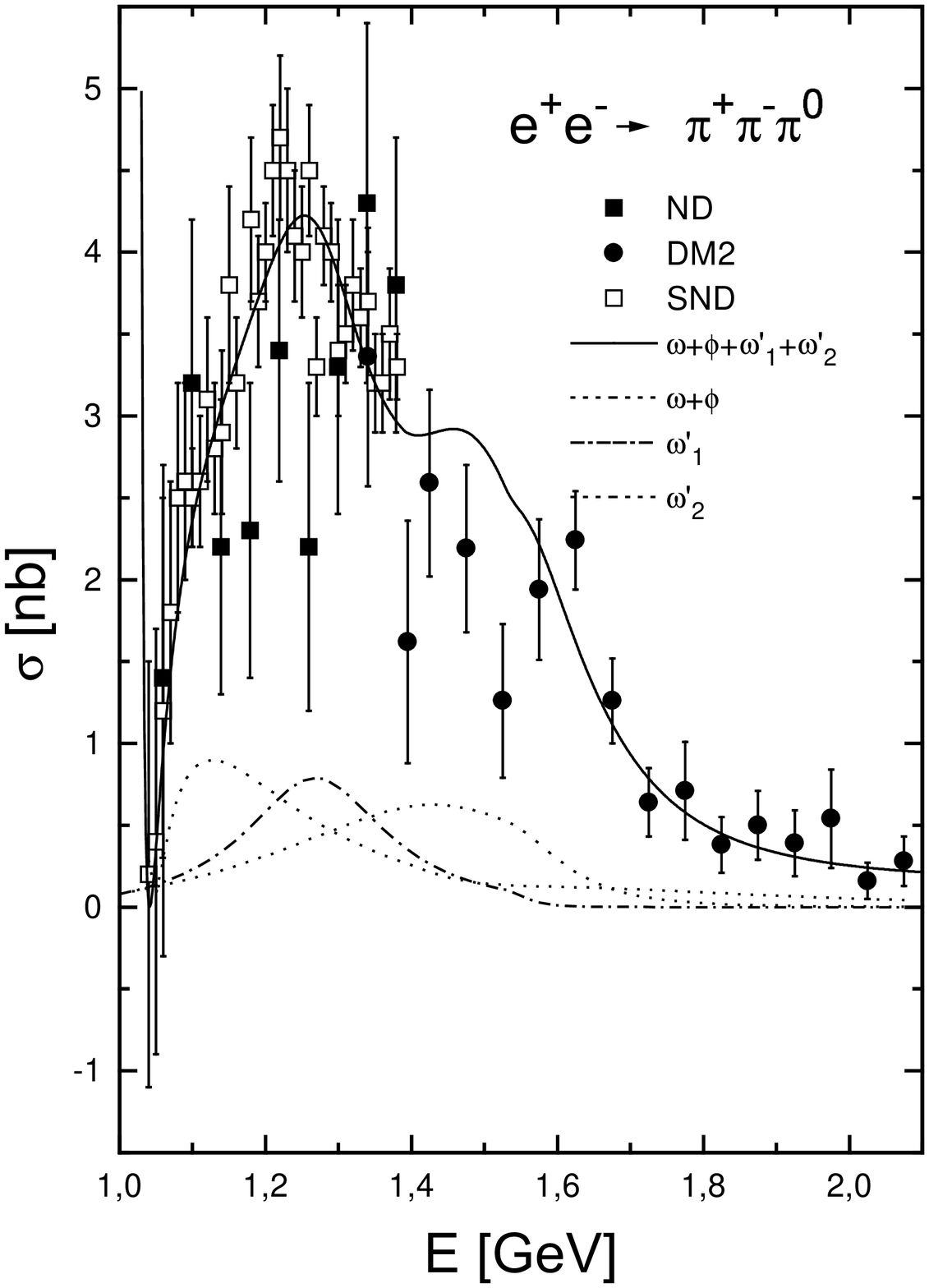}} \caption{The cross
section of the reaction $e^+e^-\to\pi^+\pi^-\pi^0$ above 1 GeV.
The data are: SND \protect\cite{snd99}, ND \protect\cite{nd91},
DM2 \protect\cite{dm2}.\label{fig1}}
\end{figure}

The similar situation is with the state $\omega(1200)$ observed
recently by the SND team in the reaction
$e^+e^-\to\pi^+\pi^-\pi^0$ \cite{snd99}. Since the rather old data
on the latter reaction were used in the fit \cite{ach98}, the new
data on this reaction \cite{snd99,cmd298} are included in the fit
done to consider the $\omega(1200)$ state in the framework of the
approach \cite{ach97,ach98}, however, upon neglecting the
contributions of the $\phi^\prime_{1,2}$ resonances, because their
couplings in the above reaction were found to be consistent with
zero \cite{ach98}. The parameters of the $\omega^\prime_{1,2}$
resonances extracted from this new fit coincide, within errors
with those reported earlier \cite{ach98}. They are used in filling
the corresponding entries in Table \ref{tab3}. The corresponding
curves are plotted in Fig.~\ref{fig1}. Our conclusion is that the
$\omega(1200)$ state observed by SND \cite{snd99} is the same
state as $\omega(1420)\equiv\omega^\prime_1$ presented in RPP
\cite{pdg}. The shift of the visible peak as compared to the bare
mass of the resonance should be attributed, as is explained above,
to the rather large width and the rapid growth of the partial
widths with the energy increase. As far as the $\omega^\prime_2$
resonance is concerned, its huge width, see Table \ref{tab2},
results, in accord with Eq.~(\ref{posita}), in the shift of the
peak to $\simeq1600$ MeV from the bare mass $\simeq1900-2000$ MeV.
The large partial widths found in the analysis \cite{ach97,ach98},
are also in the qualitative agreement with the expectations
presented in Eq.~(\ref{wid1}) and (\ref{wid2}).

Using the recent SND data \cite{snd99}, one can draw some
conclusions about the characteristics of the radial wave function
of the bound $q\bar q$ state at the origin. Since the
$\omega^\prime_1$ resonance  is usually attributed to the $2^3S_1$
state \cite{pdg,godfrey}, one finds, using Ref.~\cite{novikov},
\begin{equation}
|R_S(0)|^2=6\pi m^3_{\omega^\prime_1}/f^2_{\omega^\prime_1}=
(25\pm10)\times10^{-3}\mbox{ GeV}^3, \label{swav}
\end{equation}
to be compared to $(38\pm4)\times10^{-3}$ GeV$^3$ computed for
$\omega(782)$ meson. Within errors, both above figures agree with
each other. On the other hand, the $\omega^\prime_2$ state is
treated as the $1^3D_1$ one, hence the second derivative of the
radial wave function at the origin is appropriate \cite{novikov}:
\begin{equation}
|R^{\prime\prime}_D(0)|^2={3\pi
m^7_{\omega^\prime_2}\over25f^2_{\omega^\prime_2}}=(10\pm8)\times10^{-3}
\mbox{ GeV}^7. \label{dwav}
\end{equation}
Note that the S-wave characteristics of the $\omega^\prime_2$
resonance analogous to Eq.~(\ref{swav}) is
$|R_S(0)|^2=(35\pm23)\times10^{-3}$ GeV$^3$.

\begin{figure}
\centerline {\epsfysize=7in \epsfbox{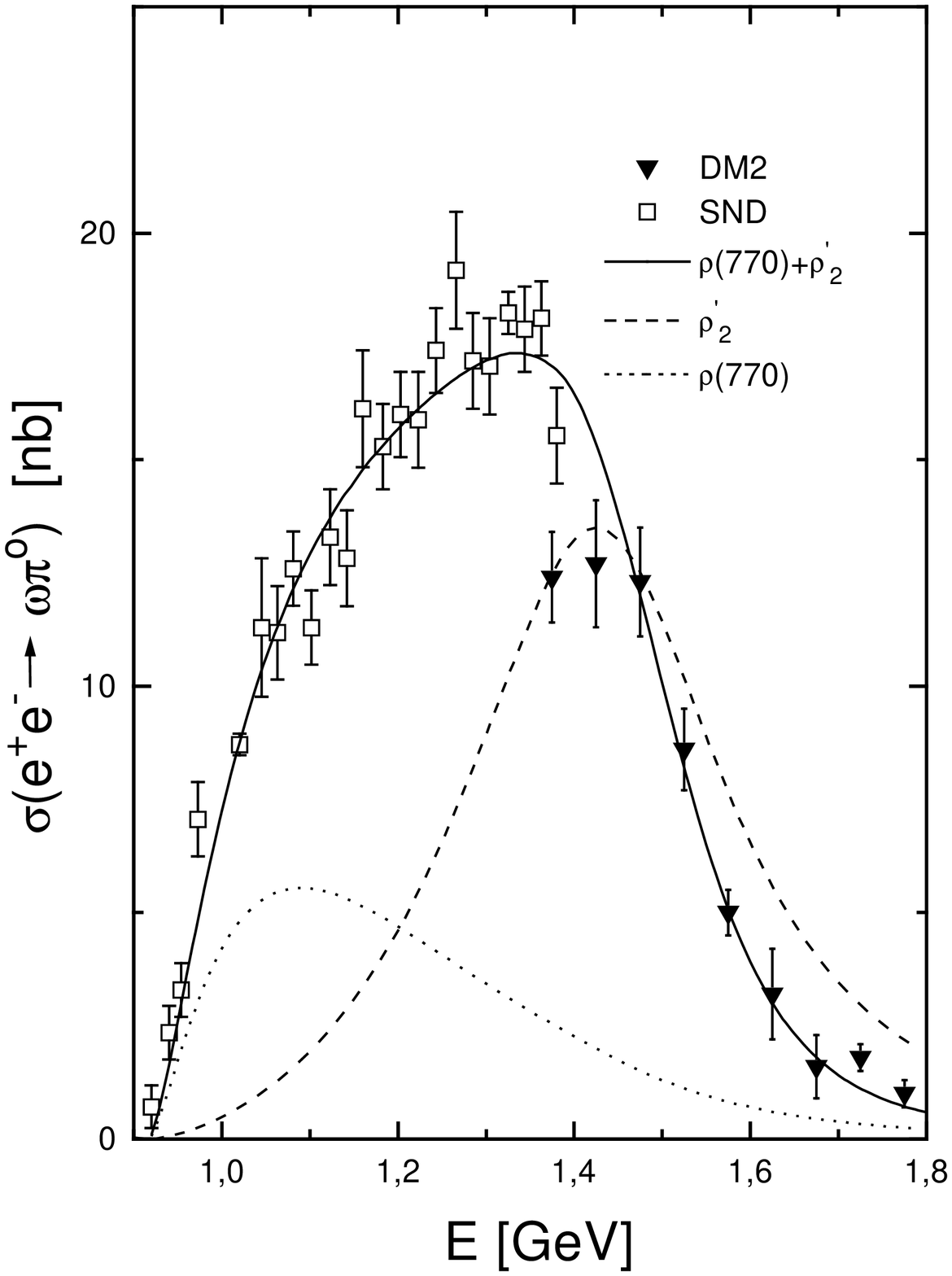}} \caption{The cross
section of the reaction $e^+e^-\to\omega\pi^0$. The data are: SND
\protect\cite{snd00}, DM2 \protect\cite{stanco}.\label{fig2}}
\end{figure}

The recent data on the reaction $e^+e^-\to\omega\pi^0$
\cite{snd00} have been included in the fit. The resonance
parameters are found to  agree within errors with those obtained
in Ref.~\cite{ach97}. Specifically, the $\rho^\prime_1$ resonance
is not revealed, since its extracted parameters are consistent
with zero within very large errors. Hence we exclude
$\rho^\prime_1$ from the fit, leaving only the
$\rho(770)+\rho^\prime_2$ contribution. Corresponding curve is
shown in Fig.~\ref{fig2}. Note that the central  value of the
range parameter $R$ entering the formfactor
\begin{equation}
C_{\rho\omega\pi}(E)={1+(Rm_\rho)^2\over1+(RE)^2}, \label{c}
\end{equation}
the latter introduced to restrict the growth of the
$\rho\to\omega\pi^0$ partial widths with the energy increase,
$\Gamma_{\rho\omega\pi}(E)\to
C_{\rho\omega\pi}^2(E)\Gamma_{\rho\omega\pi}(E)$, turns out to be
zero in our fit. To be more precise, the data allow its variation
within the interval from 0 to 0.37 GeV$^{-1}$. The difference with
the conclusion reached in Ref.~\cite{snd00} that the best fit
requires nonzero $R$ is, in our opinion, attributed to the fact
that, first,  only  two decay modes, $\pi^+\pi^-$ and
$\omega\pi^0$ were included in \cite{snd00}, while we include all
known ones (see Tables \ref{tab1} and \ref{tab2}), and the mixing
arising due to common decay modes neglected in Ref.~\cite{snd00} .
Second, the same range parameter $R$ was used for the $\pi^+\pi^-$
and $\omega\pi^0$ modes and for all the $\rho$ -like resonance,
while we have assumed it to be different, and in the case of the
$\pi^+\pi^-$ state it was set to zero, because this mode grows
rather moderately with the energy increase \cite{fn1}.

\begin{figure}
\centerline {\epsfysize=7in \epsfbox{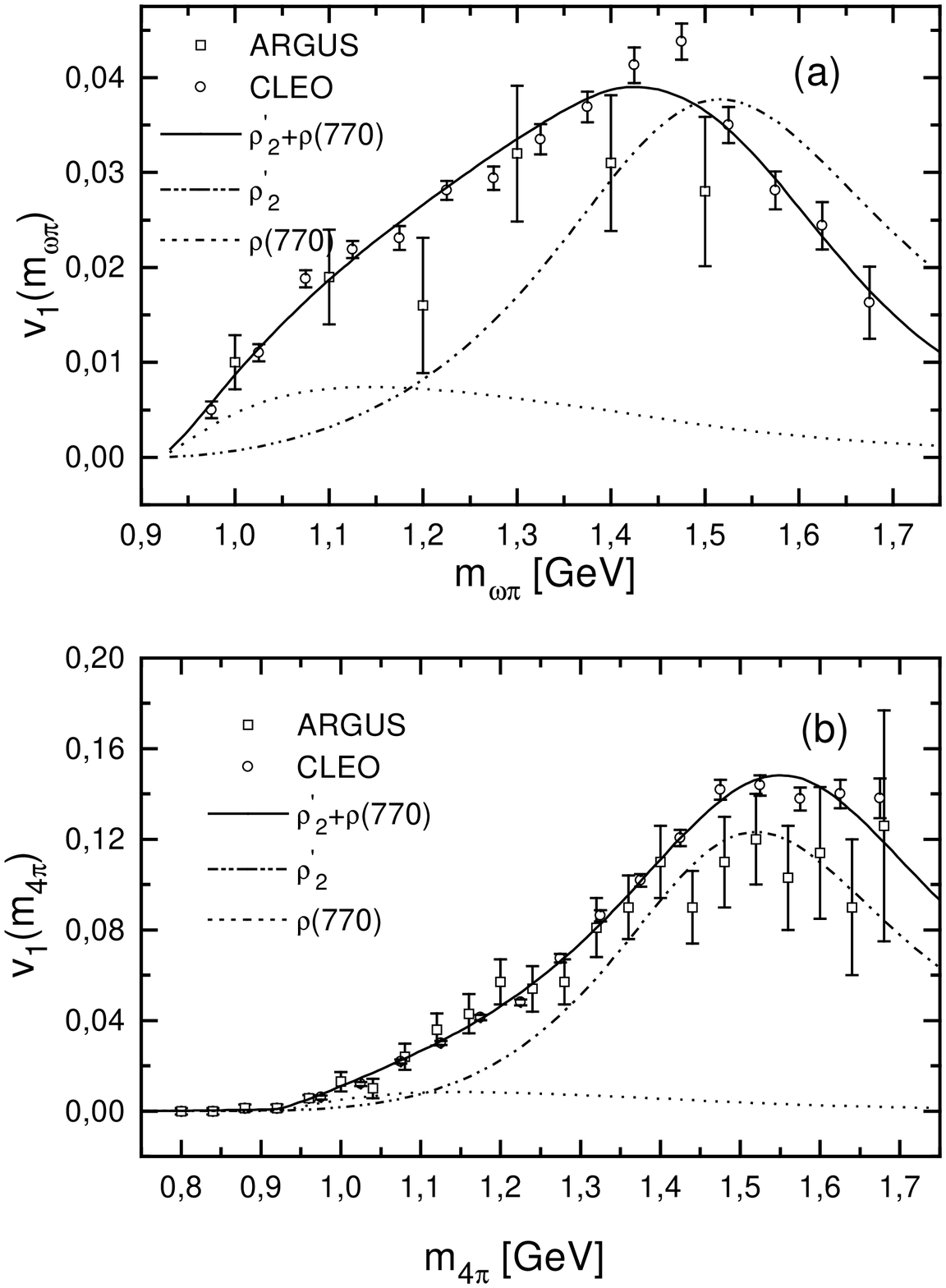}} \caption{The
spectral function in the decay $\tau^-\to\omega\pi^-\nu_\tau$ (a)
and $\tau^-\to\pi^+2\pi^-\pi^0\nu_\tau$ (b). The data are: CLEO
\protect\cite{cleo}, ARGUS \protect\cite{argus}. \label{fig3}}
\end{figure}

The joint fit of the CLEO data \cite{cleo} results in the
considerably improved accuracy of the determination of the
$\rho^\prime_2$ resonance parameters as compared to the earlier
fit \cite{ach97} of the ARGUS data \cite{argus}. Here also the
$\rho^\prime_1$ resonance is unnecessary, and the data are well
described by the only $\rho^\prime_2$ resonance in addition to the
$\rho(770)$. The results are shown in Table \ref{tab2} and
Fig.~\ref{fig3}. The shift of the visible peak towards the lower
invariant mass of the $4\pi$ system around 1500 MeV is explained
by the discussed effect exemplified by Eq.~(\ref{posita}).

Our conclusions are as follows. First,  the states
$\rho^\prime_{1,2}$ and $\omega^\prime_{1,2}$ turn out to be the
wide  resonance structures, as if the conventional quark picture
of them as the radial excitations is implied. In this respect, the
present results, having in mind their significant uncertainties,
do not contradict to the assignment of $\rho^\prime_1$ and
$\omega^\prime_1$ resonances to the state $2^3S_1$
\cite{pdg,godfrey}. In the meantime, the resonances
$\rho^\prime_2$ and $\omega^\prime_2$ are found to be wide, which
contradicts to the assigning them to the state $1^3D_1$ predicted
to be relatively narrow \cite{godfrey}. Second, one should be
careful in attributing the specific peak or structure in the cross
section to the specific spectroscopy state, because  the large
width, the rapid growth of the phase space with the energy
increase, and the mixing among the resonances  result in the shift
of the visible peaks in the cross sections. Third, the very large
widths of the resonances found in the present paper may indirectly
evidence in favor of some nonresonant contributions to the
amplitudes. The accuracy of the existing data is still poor to
isolate such contributions reliably. The forthcoming improvement
of the accuracy of the data in the energy range 1400 $-$ 2000 MeV
will hopefully permit one to specify the above nonresonant
contributions (if any) and to test the whole resonance
interpretation of the high mass states.

\acknowledgements We are grateful to V.~P.~Druzhinin,
S.~I.~Eidelman, E.~V.~Pakhtusova and S.~I.~Serednyakov for the
discussions. The present work is supported in part by the grant
RFBR-INTAS IR-97-232.

\newpage
\widetext
\begin{table}
\caption{Masses, total widths (in the units of MeV), leptonic
widths (in the units of keV), and branching ratios (percent) of
the $\rho^\prime_1$ resonance,  calculated using the coupling
constants extracted from the fits of the specific channel
\protect\cite{ach97}. The symbol $\sim$ means that the central
value is given, with the errors exceeding it considerably. The
$\rho^\prime_1$ resonance does not reveal itself in the
$e^+e^-\to\omega\pi^0$ reaction and $\tau^-$ decay.}
\begin{tabular}{ccccccc}
channel&$\pi^+\pi^-$\tablenotemark[1]&$\rho\eta$\tablenotemark[1]
&$2\pi^+2\pi^-$\tablenotemark[1]&
$\pi^+\pi^-2\pi^0$\tablenotemark[1]&$J/\psi\to3\pi$&$K^-p\to\pi^+\pi^-\Lambda$\\
\tableline
$m_{\rho^\prime_1}$&$1370^{+90}_{-70}$&$1460\pm400$&$1350\pm50$&
$1400^{+220}_{-140}$&$1570^{+250}_{-190}$&$1360^{+180}_{-160}$\\
$B_{\rho^\prime_1\to\pi^+\pi^-}$&$1.1\pm1.1$&$\sim3.7$&$\sim1.4$&
$\sim8.0$&$\sim$&$\sim0.7$\\
$B_{\rho^\prime_1\to\omega\pi^0}$&$86.5\pm41.5$&$\sim56.9$&$93.6\pm60.0$&
$77.8\pm62.2$&$66.5\pm65.5$&$93.3\pm82.7$\\
$B_{\rho^\prime_1\to\rho\eta}$\tablenotemark[2]&$\sim5.6$&$\sim3.6$&$\sim5.0$&
$\sim6.6$&$\sim13.2$&$\sim5.5$\\$B_{\rho^\prime_1\to K^\ast K\bar
K+cc}$\tablenotemark[2]&0&$\sim4.3$&0&$\sim0.4$&$\sim15.0$&0\\
$B_{\rho^\prime_1\to4\pi}$&$\sim6.8$&$\sim31.5$&$\sim0.2$&$\sim7.2$&
$\sim4.4$&$\sim0.7$\\ $\Gamma_{\rho^\prime_1\to
l^+l^-}$&$6.4^{+1.2}_{-1.4}$&$\sim13$&$5.4^{+2.6}_{-1.8}$&$6.3^{+3.3}_{-2.5}$
&$-$&$-$\\
$\Gamma_{\rho^\prime_1}$&$763\pm500$&$\sim2222$&$\sim518$&$\sim970$&$\sim3444$&$\sim460$\\
\end{tabular}
\tablenotetext[1]{This is the final state in $e^+e^-$
annihilation.} \tablenotetext[2]{Calculated assuming SU(3)
relations among the $VVP$ coupling constants.}
 \label{tab1}
\end{table}

\begin{table}
\caption{The same as in Table \protect\ref{tab1}, but in the case
of the $\rho^\prime_2$ resonance. The latter  does not reveal
itself in the $K^-p\to\pi^+\pi^-\Lambda$ reaction. }
\begin{tabular}{cccccccc}
channel&$\pi^+\pi^-$\tablenotemark[1]&$\omega\pi^0$\tablenotemark[1]&$\rho\eta$\tablenotemark[1]
&$2\pi^+2\pi^-$\tablenotemark[1]&
$\pi^+\pi^-2\pi^0$\tablenotemark[1]&$J/\psi\to3\pi$&$\tau^-\to(4\pi)^-\nu_\tau$\\
\tableline
$m_{\rho^\prime_2}$&$1900^{+170}_{-130}$&$1710\pm90$&$1910^{+1000}_{-370}$&$1851^{+270}_{-240}$&
$1790^{+110}_{-70}$&$2080^{+160}_{-900}$&$1860^{+260}_{-160}$\\
$B_{\rho^\prime_2\to\pi^+\pi^-}$&$\sim0$&$\sim0$&$\sim0.4$&$\sim1.2$&$\sim0.4$&$\sim0$&
$1.5\pm1.4$\\
$B_{\rho^\prime_2\to\omega\pi^0}$&$\sim16.7$&$22.3\pm8.0$&$\sim1.6$&$13.4\pm3.9$&$31.0\pm18.6
$&$\sim28.4$&$18.9\pm2.8$\\$B_{\rho^\prime_2\to\rho\eta}$\tablenotemark[2]&$\sim5.9$&
$6.3\pm2.0$&$\sim0.3$&$4.6\pm1.4$&$9.6\pm8.6$&$\sim11.5$&$10.3\pm2.2$\\
$B_{\rho^\prime_2\to K^\ast\bar
K+cc}$\tablenotemark[2]&$\sim8.9$&$8.7\pm2.8$&$\sim0.9$&$6.7\pm2.0$&$14.0\pm11.2$&$\sim17.8$
&$6.8\pm1.5$\\$B_{\rho^\prime_2\to4\pi}$&$\sim68.5$&$61.2\pm7.8$&$\sim96.9$&$74.0\pm32.1
$&$45.0\pm18.0$&$\sim42.2$&$62.6\pm5.0$\\$\Gamma_{\rho^\prime_2\to
l^+l^-}$&$1.8\pm1.5$&$5.2\pm1.5$&$\sim1.1$&$4.02^{+0.28}_{-0.27}$&$4.5\pm1.3$&$-$&
$9.3\pm0.6$\tablenotemark[3]
\\$\Gamma_{\rho^\prime_2}$&$\sim303.9$&$1886\pm613$&$\sim3284$&$3123\pm296$&$3151\pm1281$&
$\sim9386$&$3255\pm388$\\
\end{tabular}
\tablenotetext[1]{This is the final state in $e^+e^-$
annihilation.} \tablenotetext[2]{Calculated assuming SU(3)
relations among the $VVP$ coupling constants.}
\tablenotetext[3]{Found assuming the CVC relation between the
spectral function and the combination of the $e^+e^-$ annihilation
cross sections, see Ref.~\protect\cite{tau}.}  \label{tab2}
\end{table}
\begin{table}
\caption{Masses, total widths (in the units of MeV), leptonic
widths (in the units of eV), and branching ratios (percent) of the
$\omega^\prime_{1,2}$ resonances,  calculated using the coupling
constants extracted from the fits of the specific channel of
$e^+e^-$ annihilation \protect\cite{ach98}.  The symbol $\sim$
means that the central value is given, with the errors exceeding
it considerably.}
\begin{tabular}{cccccc}
channel&$\pi^+\pi^-\pi^0$&$\omega\pi^+\pi^-$&$K^+K^-$&$K^0_SK^\pm\pi^\mp$&$K^{\ast
0}K^\mp\pi^\pm$\\   \tableline
$m_{\omega^\prime_1}$&$1430^{+110}_{-70}$&$\sim1400$&$\sim1460$&$\sim1500$&$\sim1380$\\
$B_{\omega^\prime_1\to3\pi}$&$\sim21.6$&$\sim8$&$\sim67$&$\sim96$&$\sim34$\\
$B_{\omega^\prime_1\to K^\ast\bar
K+cc}$&$\sim0.2$&$\sim0$&$\sim1$&$\sim4$&0\\
$B_{\omega^\prime_1\to K^\ast\bar
K\pi}$&$\sim0$&0&0&0&0\\$B_{\omega^\prime_1\to\omega\pi^+\pi^-}$&$\sim78.2$&$\sim92$&
$\sim31.2$&$\sim0$&$\sim65.8$\\$\Gamma_{\omega^\prime_1\to
l^+l^-}$&$144^{+94}_{-58}$&$\sim0.2$&$\sim8$&$\sim8$&$\sim48$\\
$\Gamma_{\omega^\prime_1}$&$\sim903$&$\sim129$&$\sim173$&$\sim1252$&$\sim112$\\
$m_{\omega^\prime_2}$&$1940^{+170}_{-130}$&$2000\pm180$&$1780^{+170}_{-300}$&$\sim2120$&
$1880^{+600}_{-1000}$\\$B_{\omega^\prime_2\to3\pi}$&$\sim22.1$&$\sim34.2$&$\sim88.8$&
$\sim91.2$&$\sim60.1$\\$B_{\omega^\prime_2\to K^\ast\bar
K+cc}$&$\sim3.5$&$\sim5.8$&$\sim11.2$&$\sim15.8$&$\sim8.9$\\$B_{\omega^\prime_2\to
K^\ast\bar
K\pi}$&$\sim68.2$&$\sim53.4$&0&0&$\sim30.9$\\$B_{\omega^\prime_2\to\omega\pi^+\pi^-}$&
$\sim6.2$&$\sim6.6$&0&0&$\sim0$\\
$\Gamma_{\omega^\prime_2l^+l^-}$&$109^{+58}_{-46}$&$531\pm225$&0&$\sim189$&$1162\pm922$\\
$\Gamma_{\omega^\prime_2}$
&$\sim14000$&$\sim5757$&$\sim2420$&$\sim9854$&$\sim13820$\\
\end{tabular}
\label{tab3}
\end{table}

\begin{references}
\bibitem{fn1}
More precisely, the authors of Ref.~\cite{snd00} use the
formfactor
$C_{\rho\omega\pi}(E)=\left\{1+\left[Rq_{\omega\pi}(E)\right]^2\right\}^{-1}$
which shares with Eq.~(\ref{c}) the property that
$E\Gamma_{\rho\to\omega\pi}(E)\to const$ at $E\to\infty$.
\end{references}
\end{document}